# Using Wi-Fi Signal as Sensing Medium: Passive Radar, Channel State Information and Followups

Bo Tan, Bo Sun
Faculty of Information Technology and Communication Sciences, Tampere University, Finland
bo.tan@tuni.fi, bo.sun@tuni.fi

**Abstract**

The idea of exploiting the Wi-Fi bursts as the medium for sensing purposes, particularly for the human targets in the indoor environment, was cultivated in both radar and computer science communities and it has became a noticeable research genre with cross-disciplinary impact in security, healthcare, human-machine interaction etc.This article comparatively introduces passive radar based and channel state information (CSI) based approaches. For each means, the primary design principles, signal processing and representative applications scenarios are shown. At last, some opportunities and challenges of Wi-Fi sensing are pointed out for the sake of stepping closer to the practitioners and end-users.

**A Brief Rewind**

With the broad rollout of the Wi-Fi access points (AP) back to the middle 2000s, radar researchers started exploring the capability of sensing indoor and outdoor targets [1] with this almost free-of-charge radio resource. In the early works, the key focus of passive Wi-Fi radar are the correlation property of the waveform in time and frequency domains by using the ambiguity function and the constant false alarm rate (CFAR) algorithms proposed for the Wi-Fi waveforms. At the same time, intensive passive Wi-Fi radar experiments were conducted in the high-clutter indoor, even through-the-wall scenarios [2], to detect various types of targets. With the path paved by early works, the researchers also conducted works on optimized signal processing flows for real-time detection and variant estimation algorithms that help capture the status of the detailed motion [3]. Due to the coarse range resolution caused by the limited bandwidth of early version 802.11 signals for the indoor scenarios, most of the passive Wi-Fi radar works leverage the estimated Doppler as the primary metric because of the flexible control of the Doppler resolution. A cluster of attempts had been made on capturing the high-resolution Doppler sequences, which indicates the detailed motion status of the detected targets, especially for human targets. The prosperity of the passive Wi-Fi Doppler radar extends the scope of conventional radar to bioinformatics (e.g., activity recognition, limb movement characterization, vital signs like respiration and heartbeat detection) sensing for healthcare and facilitates a new research genre on machine learning aided radar results interpretation [4].

On the other track, the release of the open-source channel state information (CSI) exporting tools such as Linux 802.11n CSI Tool [5] and Atheros CSI Tool [6] facilitates a wave of research on terminal localization and tracking based on the angle and delay extracted from CSI. Representative work in this wave is the joint subspace angle-delay estimation based on the frequency and spatial channel samplings [7]. Besides using the CSI for direct positioning and localization of the Wi-Fi devices, another interesting trend is the use of the CSI to interpret the dynamic environment, especially the dynamic introduced by the human activities. For example, passive indoor tracking [8] and device-free gesture recognition with in-home Wi-Fi signals [9]. The recent works in this trend have revealed incredible subtle dynamics contained CSI, which can be used for key-stroking recognition, vital sign detection and bioindication. The principle of these works is in line with the passive radar if we analogize the transmitting and receiving pilot signals as the reference and surveillance channels.

From the literature, it is not difficult to notice that detecting and understanding the motion and activities of human targets is the most investigated topic because of its promising potential in healthcare, security and human-system interaction contexts. Due to the non-rigid kinematic nature of the human target, its impact on the radio includes the reflections of a collection of moving, rotating, and even vibrating body linked segments (regardless of the absorbing effect). Unlike the rigid point target, the non-rigid targets result in complicated time delays and frequency shifts representation of the reflected radio Wi-Fi signals. Though efforts have been made to build an analytical model of the human target, such as the Boulic-Thalmann model, it is still difficult to fully cover the complex kinematics of all body parts, occlusion (one body part obscuring the others) and environmental factors such as the observation perspective, multipath and interferences. Therefore, researchers leverage the data-driven means, especially the deep neural network (DNN), to resolve the problems from gesture/activity modelling to target identification for passive radar and CSI-based tracks.

In this article, we depict the premier principles of passive radar and CSI based Wi-Fi sensing and their potential variants in the recent future. We also point out concerns and challenges of Wi-Fi sensing towards practical applications.

**Passive Wi-Fi Radar**

*A. A Parasitic System*

Passive Wi-Fi radar is usually considered a parasite to the current Wi-Fi system. It does not require any change and it does not introduce any external radio signal to the existing system. Usually (depending on the deployment layout), the reference signal is expected to be received from the direct path without strong multipath impact and reflection components. The surveillance signals are expected to be received from the other paths, which potentially contain the reflections from targets of interest. The fundamental principle is to extract the differences between the reference signal and the surveillance signals. A commonly used method to jointly acquire the time and frequency domain differences is cross ambiguity function (CAF):

$$CAF(\tau, f) = \int_{t_1}^{t_2} e^{-2j\pi f t} r(t) s^*(t-\tau)\, dt \quad (1)$$

where $r(t)$ and $s(t)$ are the reference and surveillance signals, respectively, it is easy to see that dominant factors of the passive radar performance are: ***i)**,* the quality of the reference signal; ***ii)**,* the separation of the reference and surveillance channels. Thus, a normalized least mean squares (NLMS) filter or direct signal interference cleaning (DSI) algorithms must be applied before and after the CAF processing.

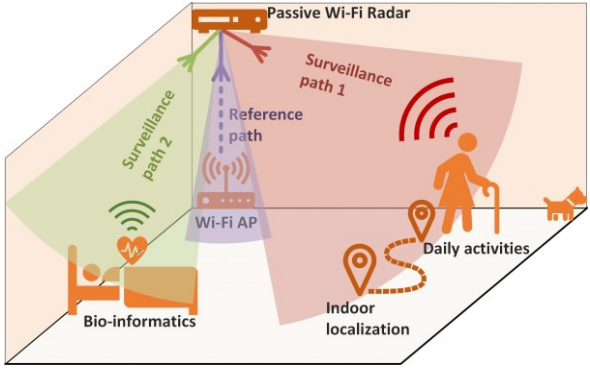

*Figure 1.* The concept schematic of Passive Wi-Fi Radar in an indoor context

*B. Micro-Doppler*
Due to historical reasons, the early 802.11 standards only defined the signal bandwidth up to 40 MHz. The signal leads to approximate 10 meters range resolution, which is almost useless for the indoor application. The Doppler becomes the primary metric used for most passive Wi-Fi radars. To some extent, the Doppler resolution is controllable by adjusting the integration duration of equation (1). Thus, by choosing the integration duration properly, passive Wi-Fi radars can be used for different human activities. Fig. 2 shows an example of the sequence of Doppler detections caused by the walking human, which includes forward and backward directions. Besides visualizing the walking, the detection of the Wi-Fi Doppler can be fine-grained to imply the subtle human body motion consists of the chest movement caused by the respiration and hand gestures. This merit makes the passive Wi-Fi radar a promising option for residential healthcare applications.

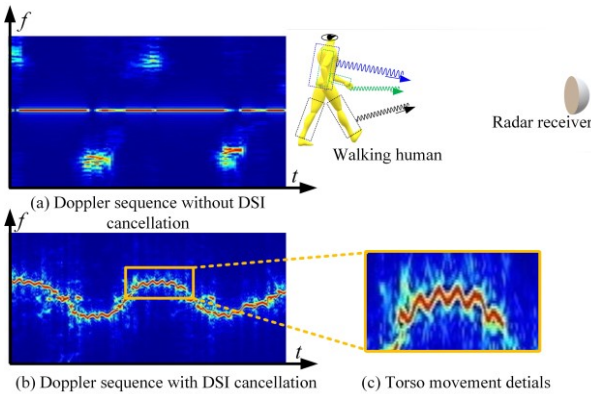

*Figure 2.* The Doppler sequence of a walking human, (a) without DSI cancellation; (b) with DSI cancellation; (c) detailed gait feature caused by torso.

*C. Reference-free Passive Radar*
The requirements of the high-quality reference signal and the separation of the reference and surveillance channels severely limit the deployment flexibility of the passive Wi-Fi radar in practice. To cast off this constraint, researchers have started to look for reference-free passive radar options. The premier approach is to take advantage of the prior knowledge of the wireless communications frames (e.g., predefined preamble sequence at the head of the frame before the data transmission) to ***i)**,* recover the original transmitting signal for CAF processing; ***ii)**,* interpertating the environment dynamic by evaluating distortion of the receiving signal. The later operation is virtually equivalent to the CSI based sensing approach.

## Sensing with CSI

*A. A Layer in Communications Receiver*
It is often called the sensing layer, that uses the predefined pilot sequences (preamble sequence in 802.11 standards) to obtain the accurate channel status, including frequency offsets, time offsets and magnitude attenuation between the transmitter and the receiver. The estimated offsets are often used for frequency tuning, time synchronization and channel equalization in the radio communication systems. However, most Wi-Fi device vendors do not open the CSI data directly to the end-user. The developers can read the CSI of some specific devices (e.g., Intel 5300, Atheros QCA9558) via the open-source drivers [5,6]. Considering the Intel 5300 as an example, it uses pilot OFDM symbols in the 802.11n signal to estimate the CSI and reports the $30 \times 3$ channel matrix (30 subcarriers from 3 receiving antennas). Each matrix consists of complex entries. Each entry can be written as $H = [|h_{ij}|e^{-j\varphi_{ij}}]$, where $h_{ij}$ and $\varphi_{ij}$ are the channel magnitude and phase of the *i*th receiving antenna and *j*th subcarrier.

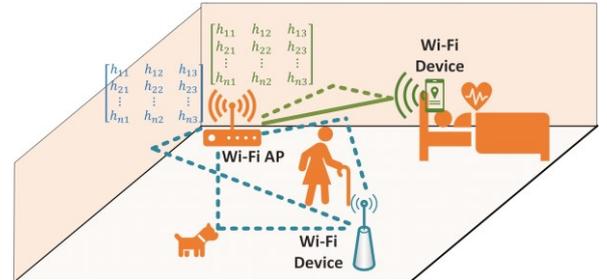

*Figure 3.* The illustration of the Wi-Fi CSI based sensing and positioning

*B. A Good Sources for Localization*
The CSI matrix ***H*** can be considered as the 2D (spatial and frequency) samples of the propagation channel, which can be used for inferring the spatial and temporal property of the propagation. If we assume the direct path dominates the propagation, the phase on each antenna (spatial domain sample) and subcarrier (frequency domain sample) provide an opportunity to localize the signal source by the joint angle-time estimation. In [7] and other works with a similar purpose, the Schmidt orthogonalization based subspace estimation and time-spatial smoothing techniques are applied for direct Wi-Fi device localization. There may not always be dominant direct components in the indoor environment, rather a collection of scattering components. In this case, the multiple carrier nature of the Wi-Fi signal provides a high degree of freedom and resolution on the temporal domain to compensate for the limited spatial resolution brought with only three antennas when using the subspace approach. With this advantage, the Wi-Fi CSI matrix *H* is also used for localizing and tracking the passive targets [8], which scatter the Wi-Fi signal to the receiver.

*C. Sensing with the Help of Deep Learning*

In practical applications, the information in need is not as straightforward as locations, for example, the body poses, gestures and the sequence of the movements needed in the residential healthcare applications. In addition, due to the kinematic complicacy of the human movement and high volume of environmental factors, it is almost impossible to build an analytic relation for specific activities and CSI measures. The recently published papers demonstrate that the DNN is an inevitable approach to handle the complex features and environmental dependence introduced by diversity of subjects, activities, perspectives, layouts, clutters characteristics. Obtaining comprehensive datasets and the transfer learning techniques become the mutual reinforcing pair to straighten out the relations between tremendous variables from subjects and environment.

**New Opportunities and Challenges**

*A. Extend Spatial Freedom on the New Spectrum*

The rollout of the new standards like 802.11 ad/ay activates the mmWave bands, bringing extra spectrum resources for the high-speed data transmissions. Intuitively, the sensing will also benefit from mmWave signal because of the finer range resolution and more sensitive Doppler shift brought by the extended bandwidth and high carrier frequency. The more important benefit of the mmWave signal may be the new level of freedom on the spatial domain, which is brought by highly dense antenna elements that can be accommodated in the limited area. Constrained by the physical size, the wide beams of the sub-6 GHz signal have a limited spatial resolution to pinpoint a target. Meanwhile, they cause more multipaths which are challenging to be resolved with coarse range and spatial resolution. Particularly for the passive radar operation, the high directivity of the mmWave signal provides high-quality reference signal, reference-surveillance separation and more surveillance channels for high-capacity passive sensing.

*B. Modal Fusion*

Though the DNN is an excellent tool to handle the high volume variables in the human target activity modelling, most of the current passive radar or Wi-Fi CSI based human sensing works are only verified in the specific environment due to the lack of the commonly accepted referencing datasets. Achieve the domain independence model has been an explicit prospect in the community. But the pace is slow due to the difficulties on large-scale data collection in various environments and annotation of the massive volumes of data. Therefore, having the passive Wi-Fi radar or CSI dataset accompany the other data modals (e.g., video, acoustics and infrared) as ground truth or cross-validation will be a path to facilitate the data resources growth. The OPERAnet [10], which contains the passive radar, CSI, and UBW and depth sensor, is a valuable pilot towards this direction. RF-pose [11] proves the feasibility of cross-modal training for radio sensing data.

*C. Privacy: a Double-Edged Sword*

Most Wi-Fi signal based sensing propositions claims privacy-preserving as the one significant benefit compared with the video solutions, which take the identifiable image information of the user. It becomes an antinomy when the Wi-Fi becomes more and more accurate because it becomes private information. In this context, the privacy concern will be the main impedance to the popularity of the Wi-Fi sensing concept. Because of the tradition, even physical layer security solutions will be invalid as the sensing system often uses radio frequency preparties. Thus, the new privacy-preserving paradigm from the waveform level [12] is on the agenda to pave the path for Wi-Fi sensing.

**Conclusion**

The past works from researchers of the radar, electronics and computer science communities have proved the feasibility and illustrated the prospect of sensing within the Wi-Fi framework. However, as an underlying service in the standardized communications system, the quality of the service (QoS) needs to meet a certain referable standard. Thus, it still needs tremendous efforts of the community to build the standard reference performance metric in typical scenarios. Furthermore, we also need catalytic factors such as the system on a chip (SoC) solution of the sensing function, innovative user-centric service design trustable privacy-preserving mechanism to finally bring the Wi-Fi sensing (regardless of the forms) to the end-users from the laboratory.

**References**

[1] F. Colone, P. Falcone, C. Bongioanni and P. Lombardo, "WiFi-Based Passive Bistatic Radar: Data Processing Schemes and Experimental Results," in IEEE Transactions on Aerospace and Electronic Systems, vol. 48, no. 2, pp. 1061-1079, April 2012

[2] K. Chetty, G. E. Smith and K. Woodbridge, "Through-the-Wall Sensing of Personnel Using Passive Bistatic Wi-Fi Radar at Standoff Distances," in IEEE Transactions on Geoscience and Remote Sensing, vol. 50, no. 4, pp. 1218-1226, April 2012

[3] B. Tan, K. Woodbridge and K. Chetty, "A real-time high resolution passive Wi-Fi Doppler-radar and its applications," International Radar Conference, Sept 2014

[4] B. Tan, Q. Chen, K. Chetty, K. Woodbridge, W. Li and R. Piechocki, "Exploiting Wi-Fi Channel State Information for Residential Healthcare Informatics," in IEEE Communications Magazine, vol. 56, no. 5, pp. 130-137, May 2018

[5] D. Halperin, W. Hu, A. Sheth, D. Wetherall, "Tool Release: Gathering 802.11n Traces with Channel State Information", ACM SIGCOMM CCR, vol. 48, no. 2, pp. 53, Jan 2011.

[6] Y. Xie, Z. Li, M. Li. 2015. "Precise Power Delay Profiling with Commodity Wi-Fi" In Proceedings of the 21st Annual International Conference on Mobile Computing and Networking (MobiCom). ACM, New York, NY, USA, 53-64. 2015

[7] M. Kotaru, K. Joshi, D. Bharadia, S. Katti.. "SpotFi: Decimeter Level Localization Using Wi-Fi" SIGCOMM Comput. Commun. Rev. 45, 4, 269–282, October 2015

[8] Y. Xie, J. Xiong, M. Li, K. Jamieson. "MD-Track: Leveraging Multi-Dimensionality for Passive Indoor Wi-Fi Tracking" In The 25th Annual International Conference on Mobile Computing and

Networking. (MobiCom), ACM, New York, NY, USA, Article 8, 1–16, 2019

[9] Q. Pu, S. Gupta, S. Gollakota, S. Patel. "Whole-home gesture recognition using wireless signals" In Proceedings of the 19th annual international conference on Mobile computing & networking (MobiCom). ACM, New York, NY, USA, 27–38, 2013

[10] M. J. Bocus, W. Li, S. Vishwakarma, R. Kou, C. Tang, K. Woodbridge, I. Craddock, R. McConville, R. Santos-Rodriguez, K. Chetty, R. Piechocki. "OPERAnet: A Multimodal Activity Recognition Dataset Acquired from Radio Frequency and Vision-based Sensors", arXiv:2110.04239, Oct, 2021

[11] M. Zhao et al., "Through-Wall Human Pose Estimation Using Radio Signals," IEEE/CVF Conference on Computer Vision and Pattern Recognition (CVPR), pp. 7356-7365, 2018

[12] T. Xu "Waveform-Defined Privacy: A Signal Solution to Protect Wireless Sensing", IEEE 94th Vehicular Technology Conference, 2021

## Biography

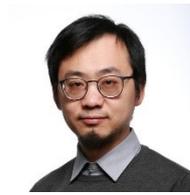

**Bo Tan** is a tenure track assistant professor in Tampere University, Finland. His research interests include radio signal processing in radar, wireless communications systems, joint sensing communications design, machine learning of multiple modal sensing data fusion for human activity recognition, privacy preserving and security mechanism of sensing data. He is principal investigator of multiple EU and Finland research projects. He is also an active reviewer of IEEE and IET journals in radio techniques.

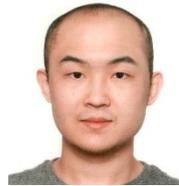

**Bo Sun** completed his Master of Science (MSc) in the field of Wireless Communication and RF from Tampere University in 2019. He then joined the Faculty of Information Technology and Communication Sciences, Tampere University, to continue studying to Ph.D. His research interests include 5G communication systems, OFDM, radar systems, and reconfigurable intelligent surface.